\documentclass[floatfix,showpacs,prl,twocolumn]{revtex4}
\usepackage{makeidx}
\usepackage{eurosym}
\usepackage{graphicx}
\usepackage{graphics}
\usepackage{amssymb}
\usepackage{amsmath}
\usepackage{color}

\begin{document}

\title{Self-consistent Bogoliubov de Gennes theory of the vortex lattice
state in a two-dimensional strong type-II superconductor at high magnetic
fields }
\author{Vladimir Zhuravlev}
\affiliation{Schulich Faculty of Chemistry, Technion-Israel Institute of Technology,
Haifa 32000, Israel }
\author{Wenye Duan}
\affiliation{School of Physics, Peking University, Beijing 100871, China }
\author{Tsofar Maniv}
\affiliation{Schulich Faculty of Chemistry, Technion-Israel Institute of Technology,
Haifa 32000, Israel}
\email{e-mail:maniv@tx.technion.ac.il}
\date{\today }

\begin{abstract}
A self-consistent Bogoliubov deGennes theory of the vortex lattice state in
a 2D strong type-II superconductor at high magnetic fields reveals a novel
quantum mixed state around the semiclassical H$_{c2}$, characterized by a
well-defined Landau--Bloch band structure in the quasi-particle spectrum and
suppressed order-parameter amplitude, which sharply crossover into the
well-known semiclassical (Helfand-Werthamer) results upon decresing magnetic
field. Application to the 2D superconducting state observed recently on the
surface of the topological insulator Sb$_{2}$Te$_{3}$, accounts well for the
experimental data, revealing a strong type-II superconductor, with unusualy
low carrier density and very small cyclotron mass, which can be realized
only in the strong coupling superconductor limit.
\end{abstract}

\maketitle

{The recent discoveries of surface and interface superconductivity with} {%
exceptionally high superconducting (SC) transition temperatures in several
material structures} \cite{BosovicNphys14},\cite{FengGeNmat15}, \cite%
{MannaArXive16} {promise to stimulate fundamental studies of the phenomenon
of strong type-II superconductivity in two-dimensional (2D) and quasi 2D
electron systems, particularly under high magnetic fields}.\cite{Maniv01} In
a pure strong type-II superconductor under a uniform magnetic field the
quasi particle spectrum is gapless in a broad field range below the upper
critical field $H_{c2}$ \cite{Dukan94},\cite{Dukan95},\cite{Maniv01}, where
scattering of quasi particles by the vortex lattice interferes with the
Landau quantization of the electron motion perpendicular to the magnetic
field to form magnetic (Landau) Bloch's bands. In pure 2D, or quasi 2D,
strong type-II SC systems, such as that realized in the multilayer system of
the organic charge transfer salt $\kappa -\left( ET\right) _{2}Cu\left(
SCN\right) _{2}$ \cite{Wosnitza96}, under a magnetic field perpendicular to
the easy conducting plane, the underlying normal electron spectrum is fully
quantized and the effect of the vortex lattice is very pronounced. 2D vortex
lattices can realize in such strongly layered electronic systems due to the
presence of weak crystalline disorder \cite{Wehr06},\cite{Suderow14}, where
pinning of a few flux lines provide support for the entire vortex lattice
against melting under an increasing magnetic field up to the irreversibility
line \cite{Maniv15}. \ {Of special interest in the present paper is the
unique situation of the 2D superconductivity realized in surface states of
topological insulators, e.g. }Sb$_{2}$Te$_{3}$ \cite{Zhao15}, {where the
chemical potential} $\mu $ {is close to a Dirac point }\cite{Zhang09}{\
(with Fermi velocity }$v${) and the cyclotron effective mass}, $m^{\ast
}=\mu /v^{2}$\cite{Katsnelson12} {is a small fraction (e.g. 0.065 in }Sb$_{2}
$Te$_{3}${) of the free electron mass} $m_{e}$, {resulting in a dramatic
enhancement of the cyclotron frequency}, $\omega _{c}=eH/m^{\ast }c$, {and
the corresponding Landau level (LL) energy spacing.}

Due to the suppressed energy dispersion along the magnetic field direction,
characterizing the 2D electron system, and the particle-hole symmetry
inherent to the SC state, the quasi particle spectrum exhibits peculiar
features that are missing in the 3D case. For example, at discrete magnetic
field values where the chemical potential is located in the middle of a
Landau band, so that the underlying normal state spectrum satisfies
particle-hole symmetry, the calculated quasi-particle density of states
shows a linear, Dirac-like energy dependence, which reflects topological
singularities at the vortex lattice cores \cite{Maniv01},\cite%
{Tesanovic-Sacramento98}. \ Both the enhanced Landau quantization effect and
the lucid reflection of the topological singularity at the vortex-lattice
cores in the quasi-particle spectrum,\cite{GGLex},\cite{MZarXiv14}, point to
the great importance of self consistency in the theoretical description of
2D superconductors at high magnetic fields. The presence of well defined
Landau bands in the quasi-particle spectrum, which is reflected as {%
magneto-quantum (MQ)} oscillations in the SC order parameter, \cite{Maniv92}
via the self consistency equation, is expected to significantly alter the
semiclassical picture of the SC phase transition at high magnetic fields.%
\cite{HW64} \ In this paper we present, for the first time, results of
systematic self-consistent solutions of the Bogoliubove de Gennes (BdG)
equations, which addresses these aspects of 2D superconductivity at high
magnetic fields. It is, indeed, found that, due to self-consistency, the SC
pair-potential can be strongly distorted near the vortex-lattice cores, and
that the traditional semiclassical picture of a single critical point can be
dramatically smeared into an intermediate state of multicritical transition.
\ The self-consistency formalism used here was first developed in Ref.\cite%
{Norman95}; however, its few past applications \cite{Norman95},\cite{Kita02}
have not addressed the above mentioned aspects. \ \ For the present analysis
we consider a model of a 2D electron system under a perpendicular uniform
magnetic field $\mathbf{H=}\left( 0,0,H\right) $, neglecting, for the sake
of simplicity, any (Zeeman or spin-orbit induced) spin splitting and
assuming a singlet, $s$-wave electron pairing. The large enhancement of the
cyclotron energy in the physical model systems under consideration here \cite%
{Zhao15} justifies this approximation (see later).

\begin{figure}[tbp]
\label{fig1a} \includegraphics[width=3.4in]{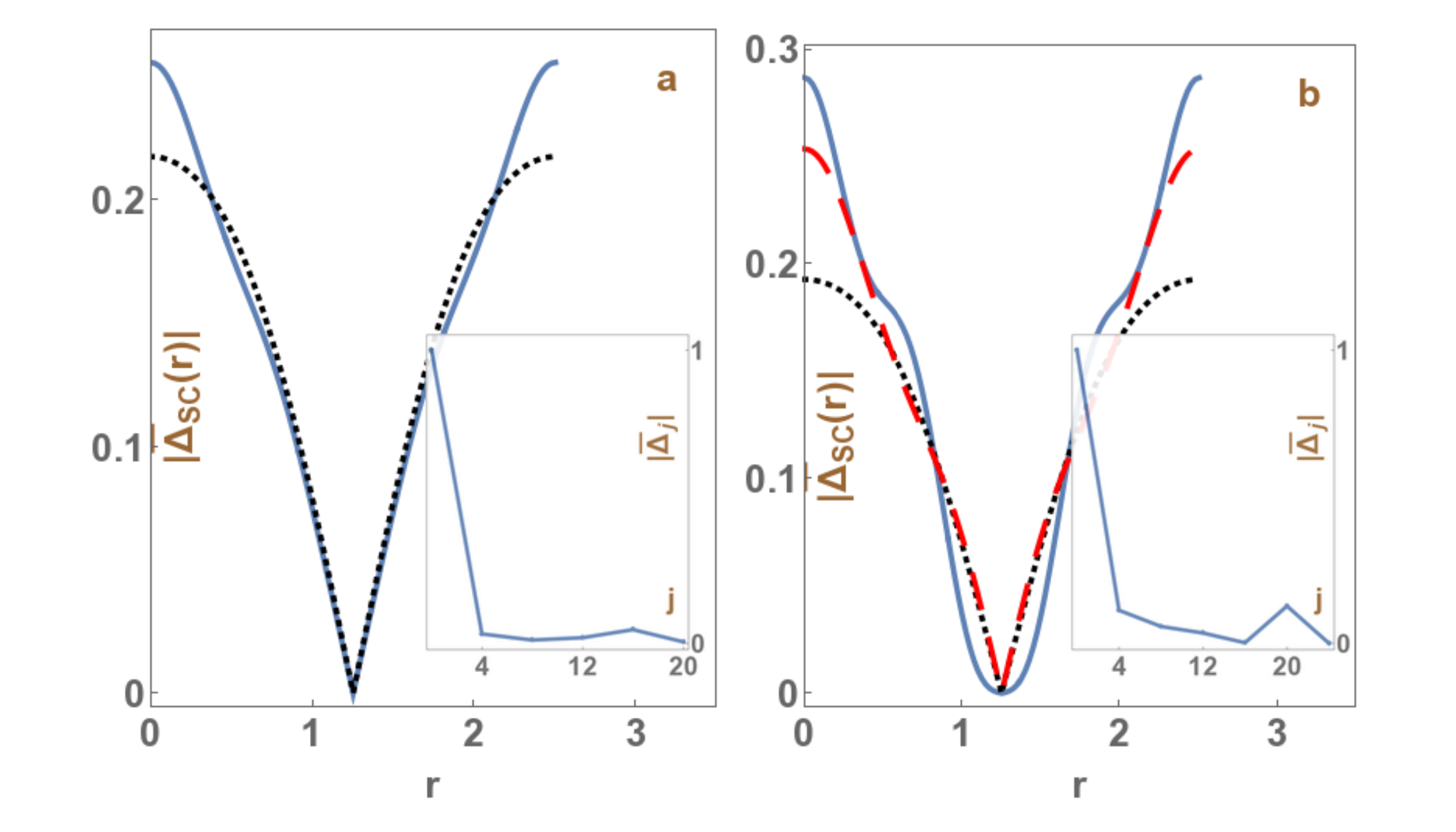}
\caption{Color online: (a) Self-consistent $\left \vert \overline{\Delta }%
_{SC}\left( \mathbf{r}\right) \right \vert $ (i.e. measured in units of $%
\hbar \protect\omega _{c}$), as a function of coordinate along the diagonal
of the unit cell in the (square) vortex lattice (blue solid line), for $%
n_{F}=9$. \ The other parameters used in the calculation are: $\protect%
\lambda =0.75,\hbar \protect\omega _{D}/\protect\mu =0.2$, and $k_{B}T/%
\protect\mu =10^{-4}$. \ For comparison, the resulting $\left \vert 
\overline{\Delta }_{SC}\left( \mathbf{r}\right) \right \vert $\ calculated
only with $\overline{\Delta }_{j=0}$, which corresponds to the Abrikosov
solution, is also shown (dotted line). The inset shows all the nonzero
coefficients $\left \vert \overline{\Delta }_{j}\right \vert $, $%
j=0,4,8,12,16\leq 2n_{F}$. (b) Similar to (a) but for $n_{F}=10$. \ The
inset shows the nonzero coefficients $\left \vert \overline{\Delta }%
_{j}\right \vert $ at $j=0,4,8,12,16,20$ $\leq 2n_{F}$ . Note the parabolic
distortion of $\left \vert \overline{\Delta }_{SC}\left( \mathbf{r}\right)
\right \vert $ in the vortex core region and the enhanced value of $%
\left
\vert \overline{\Delta }_{j}\right \vert $ at $j=2n_{F}.$ For
comparison, the resulting $\left \vert \overline{\Delta }_{SC}\left( \mathbf{%
r}\right) \right \vert ,$\ calculated without $\overline{\Delta }_{j=2n_{F}}$%
, is also included (red dashed line), showing no distortion of the vortex
core region.}
\end{figure}

The corresponding equations for the quasi particle states in the mean-field
approximation are the BdG equations in the Landau-orbitals representation,
using the magnetic Bloch basis-set wavefunctions \cite{Dukan94,Norman95}: 
\begin{eqnarray}
\sum\limits_{n^{\prime }}\Delta _{n,n^{\prime }}\left( \mathbf{q}\right)
v_{n^{\prime }}^{N}\left( \mathbf{q}\right)  &=&\left( \varepsilon
^{N}\left( \mathbf{q}\right) -\xi _{n}\right) u_{n}^{N}\left( \mathbf{q}%
\right) ,  \notag \\
\sum\limits_{n^{\prime }}\Delta _{n^{\prime },n}^{\ast }\left( \mathbf{q}%
\right) u_{n^{\prime }}^{N}\left( \mathbf{q}\right)  &=&\left( \varepsilon
^{N}\left( \mathbf{q}\right) +\xi _{n}\right) v_{n}^{N}\left( \mathbf{q}%
\right)   \label{BdG}
\end{eqnarray}%
where the single-electron (LL) energy measured relative to the chemical
potential $\mu $ is given by $\xi _{n}=\hbar \omega _{c}\left(
n-n_{F}\right) ,n=0,1,2,...$, $n_{F}=\mu /\hbar \omega _{c}-1/2$.\ The
matrix elements,$\Delta _{nn^{\prime }}\left( \mathbf{q}\right) $, of the
self-consistent pair potential, $\Delta \left( \mathbf{r}\right) =\left\vert
V\right\vert \sum_{N}\sum_{\mathbf{q}}u_{\mathbf{q}}^{N}\left( \mathbf{r}%
\right) v_{\mathbf{q}}^{N\ast }\left( \mathbf{r}\right) \tanh \left(
\varepsilon ^{N}\left( \mathbf{q}\right) /2k_{B}T\right) $ \cite{Maniv01},
are calculated by exploiting an expansion \cite{Eilenberger67},\cite{Dukan94}%
, $\Delta \left( \mathbf{r}\right) =\sum\limits_{j=0}^{\infty }\Delta
_{j}\eta _{j}\left( \mathbf{r}\right) $, in terms of Landau orbitals
wavefunctions of a Cooper-pair (charge $2e$), $\eta _{j}\left( \mathbf{r}%
\right) =\sum\limits_{k}e^{i\pi \left( \frac{b_{x}}{a_{x}}\right) k^{2}}e^{i%
\frac{2\pi k}{a_{x}}x}\varphi _{j}\left[ \sqrt{2}\left( y+\frac{\pi k}{a_{x}}%
\right) \right] $, where $\varphi _{j}\left( y\right) \equiv \left( \sqrt{%
\frac{\pi }{2}}2^{j}j!\right) ^{-1/2}e^{-\frac{1}{2}y^{2}}H_{j}\left(
y\right) $, and $H_{j}\left( y\right) $ is hermite polynomial of order $j$.
\ Here $\mathbf{a}=\left( a_{x},0\right) ,\mathbf{b}=\left(
b_{x},b_{y}\right) $ are two primitive vectors in a general rhombic vortex
lattice, forming a primitive unit cell of area $a_{x}b_{y}=\pi $,
corresponding to one Cooper-pair flux quantum (note that all spatial length
are measured in units of the magnetic length: $a_{H}\equiv \sqrt{\frac{%
c\hbar }{eH}}$). \ Self-consistency is therefore established by requiring
the coefficients $\Delta _{j}$ to satisfy the equations: 
\begin{eqnarray}
\Delta _{j} &=&\hbar \omega _{c}\left( \sqrt{2}\pi /a_{x}\right) \left(
j!2^{j}\right) \lambda \sum_{n,m=n_{F}-n_{0}}^{n_{F}+n_{0}}\Gamma _{n,m}^{j}
\notag \\
&\times &N_{\phi }^{-1}\sum_{\mathbf{q}}\Phi _{nmj}\left( \mathbf{q}\right)
\Psi _{n,m}\left( \mathbf{q}\right)   \label{Del_j}
\end{eqnarray}%
where: 
\begin{eqnarray}
\Gamma _{n,m}^{j} &\equiv &2^{-n-m}\left( n!m!\right) ^{-1/2}\left(
-1\right) ^{m}\sum\limits_{k=0}^{j}\left( -1\right) ^{k}C_{k}^{m}C_{j-k}^{n},
\notag \\
\Psi _{n,m}\left( \mathbf{q}\right)  &\equiv &\sum_{N}u_{n}^{N}\left( 
\mathbf{q}\right) v_{m}^{N\ast }\left( \mathbf{q}\right) \tanh \left(
\varepsilon ^{N}\left( \mathbf{q}\right) /2k_{B}T\right) ,  \notag \\
\Phi _{nmj}\left( \mathbf{q}\right)  &=&\left( \sqrt{\frac{\pi }{2}}%
2^{j}j!\right) ^{-1/2}\sum\limits_{l}H_{n+m-j}\left[ \sqrt{2}\left( q_{x}+%
\frac{\pi l}{a_{x}}\right) \right]   \notag \\
&\times &e^{i\pi \left( \frac{b_{x}}{a_{x}}\right) l^{2}-i\left( \frac{2\pi l%
}{a_{x}}\right) q_{y}-\left( q_{x}+\frac{\pi l}{a_{x}}\right) ^{2}},  \notag
\end{eqnarray}%
$n_{0}\equiv \left[ \omega _{D}/\omega _{c}\right] $, $N_{\phi }$ is the
number of flux lines threading the SC sample, and $\omega _{D}$- the (Debye)
cut-off frequency, and the matrix elements: $\Delta _{nm}\left( \mathbf{q}%
\right) =\sum\limits_{j=0}^{\infty }\Delta _{j,nm}\left( \mathbf{q}\right) $%
, obeying the equations: 
\begin{eqnarray}
&&\Delta _{j,nm}\left( \mathbf{q}\right) =\Delta _{j}\left( 2\pi \right)
^{-1/4}\left( j!2^{j}\right) ^{1/2}\Gamma _{n,m}^{j}\sum\limits_{l}
\label{Del_jnm} \\
&&e^{-\frac{i\pi }{4}\left( \frac{b_{x}}{a_{x}}\right) l^{2}+i\frac{2\pi l}{%
a_{x}}q_{y}-\left( q_{x}+\frac{\pi l}{a_{x}}\right) ^{2}}H_{n+m-j}\left[ 
\sqrt{2}\left( q_{x}+\frac{\pi l}{a_{x}}\right) \right]   \notag
\end{eqnarray}%
Note the dimensionless coupling constant $\lambda $ in Eq.\ref{Del_j}, which
is related to the effective electron-electron interaction parameter $%
\left\vert V\right\vert $ , through: $\lambda \equiv \left\vert V\right\vert
\left( m^{\ast }/2\pi \hbar ^{2}\right) =\left\vert V\right\vert /2\pi
a_{H}^{2}\hbar \omega _{c}$. Note also that the range of summation over the
wavevectors $\mathbf{q}$ in Eq.\ref{Del_j} is restricted to the first
magnetic Brillouin zone, which can be further reduced to a quarter, or
one-sixth part of the zone, depending on whether the point symmetry of the
vortex lattice is four-fold or six-fold, respectively. \ For the sake of
simplicity, all numerical calculations have been performed here with the
square-lattice geometry. The small changes associated with a different
lattice geometry, e.g. the more common triangular geometry, are irrelevant
to the purposes of the present paper.

The self-consistent solution code starts with initial values of the
coefficients $\Delta _{j}$ in the expression for the matrix elements in Eq.%
\ref{Del_jnm}, for which the calculated $4\times \left( 2n_{0}+1\right)
\times \left( 2n_{0}+1\right) $ BdG matrix in the $2\times \left(
2n_{0}+1\right) $-dimensional vector space $\left( u_{n}^{N}\left( \mathbf{q}%
\right) ,v_{m}^{N}\left( \mathbf{q}\right) \right) $ is numerically
diagonalized at each point $\mathbf{q}$ in the magnetic Brillouin zone. The
resulting eigenvectors and eigenvalues are then used to construct the pair
potential for the next iteration with the new $\Delta _{j}$ and $\Delta
_{nm}\left( \mathbf{q}\right) $ through Eq.\ref{Del_j} and \ref{Del_jnm}
respectively. \ The iteration process continues until the values of $\Delta
_{j}$ converge. Note that, due to the rotational symmetry of the vortex
lattice, only $\Delta _{j}$ with $j$ integer multiple of $4$ ($6$) for the
square (triangular) lattice are different from zero. The resulting values of 
$\Delta _{j}$ for $j>0$ , in the high magnetic fields region of well-defined
Landau bands, are usually much smaller than $\Delta _{j=0}$ , which
corresponds to Abrikosov-lattice form of $\Delta \left( \mathbf{r}\right) $,
but become increasingly important upon decreasing the field below the
crossover to a continuous spectrum, where the the pair-potential shows a
reduced vortex core region and an oscillating behavior away from the core,
which reflect the appearance of bound states.\cite{Norman95} Note, however,
that at special small values of the LL filling factor $n_{F}+1/2$ ($\propto
1/H$ ), $\Delta _{j=2n_{F}}$ is strongly enhanced with respect to the other
coefficients with $j>0$, leading to a significant distortion of $\Delta
\left( \mathbf{r}\right) $ in the vortex core region, with respect to the
linearly vanishing Abrikosov form (see Fig.(1)). The excess kinetic energy
associated with this distortion is compensated by the extra condensation
energy involved in the orbital extension over the reciprocal vortex lattice.

Self-consistent determination of the pair-potential amplitude, $\Delta _{SC}$%
, is also of a crucial importance in the range of relatively small $n_{F}$
(high field) values, where {MQ} oscillations significantly influence the
transition to the SC state. \ Fig.(2) shows $\overline{\Delta }_{SC}\left(
n_{F}\right) \equiv \Delta _{SC}\left( n_{F}\right) /\hbar \omega _{c}$
(calculated at integer values of $n_{F}$, where $\Delta _{SC}\left(
n_{F}\right) $ has maxima) for the coupling constant $\lambda =0.75$ and
various values of the cut-off frequency parameter $\hbar \omega _{D}/\mu $.
The corresponding best fits of $\overline{\Delta }_{SC}\left( n_{F}\right) $
to a phenomenological Ginzburg-Landau (GL)-like formula:%
\begin{equation}
\overline{\Delta }_{GL}\left( n_{F}\right) \equiv \frac{\Delta _{GL}\left(
n_{F}\right) }{\hbar \omega _{c}}=\frac{\gamma n_{F}}{n_{c2}}\left[ 1-\frac{%
n_{c2}}{n_{F}}\right] ^{1/2}  \label{Del_GL}
\end{equation}%
with two adjustable parameters, $\gamma $ and $n_{c2}$, are also plotted in
Fig.(2). 
\begin{figure}[tbp]
\label{fig2} \includegraphics[width=3.3in]{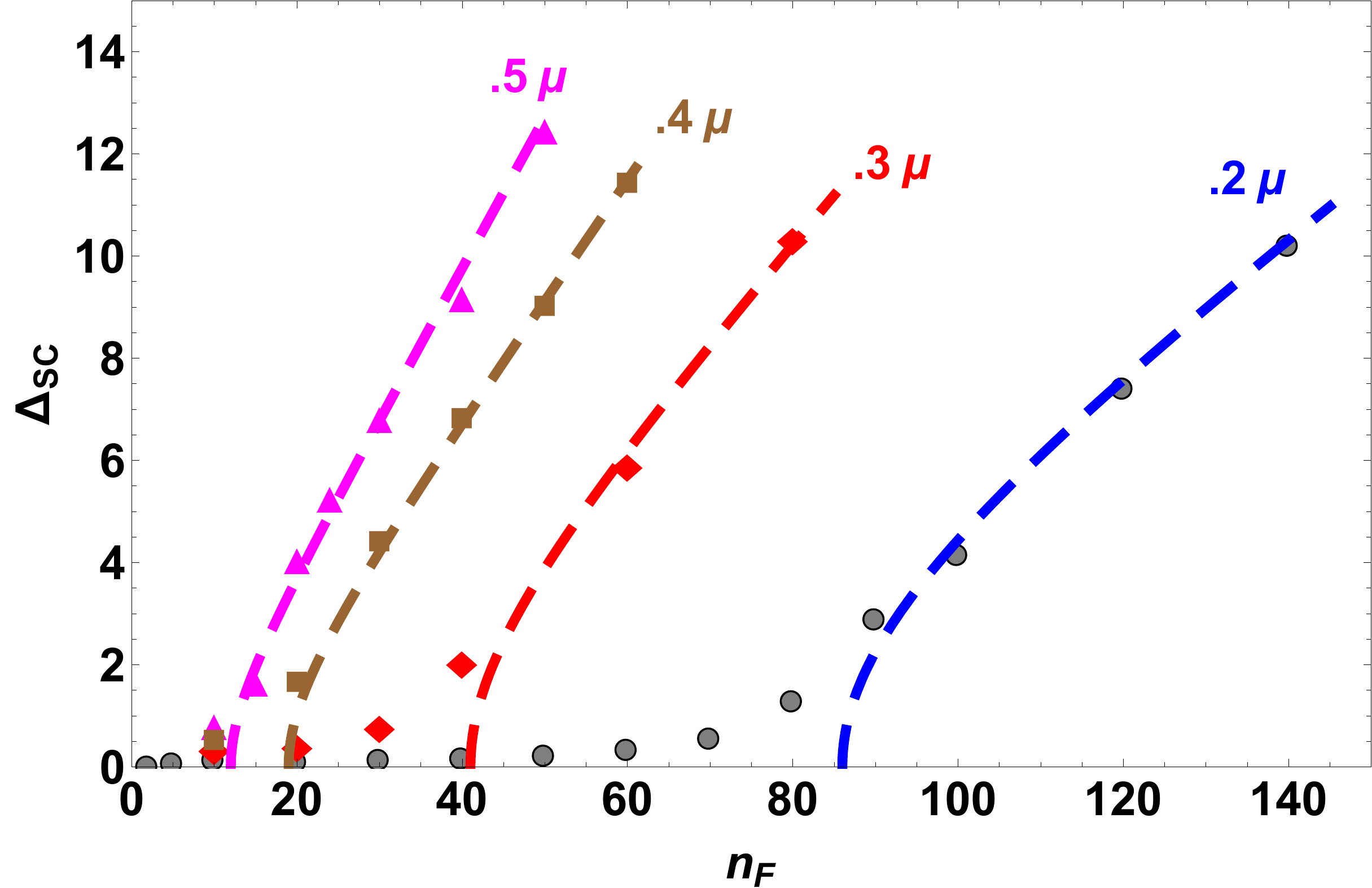}
\caption{Color online: Self-consistent order-parameter amplitde, $\overline{%
\Delta }_{SC}\left( n_{F}\right) $, (i.e. measured in units of $\hbar 
\protect\omega _{c}$), calculated at integer values of $n_{F}$, for $\protect%
\lambda =0.75$ and different values of $\protect\omega _{D}$ ($0.2\protect%
\mu $- circles, $0.3\protect\mu $-diamonds, $0.4\protect\mu $-squares, and $%
0.5\protect\mu $-triangles). The corresponding best fitting curves to $%
\overline{\Delta }_{SC}\left( n_{F}\right) $, based on the GL-like
expression 
\index{Del\_GL} are also shown (dashed lines).}
\end{figure}

The best fitting values of $\gamma $ and $n_{c2}$ are found to be close to $%
2\gamma _{HW}$ and $2n_{c2}^{HW}$ respectively, where $\gamma _{HW}\approx
0.56\left( \mu /\hbar \omega _{D}\right) \sinh \left( 1/\lambda \right) $,
and $n_{c2}^{HW}=\mu /\left( \hbar eH_{c2}^{HW}/m^{\ast }c\right)
=0.563\left( \mu /\hbar \omega _{D}\right) ^{2}\sinh ^{2}\left( 1/\lambda
\right) $, are the respective expressions derived within the semiclassical
Helfand-Werthamer (HW) theory \cite{HW64} for the order-parameter amplitude $%
\overline{\Delta }_{HW}\left( n_{F}\right) =\left( \gamma
_{HW}/n_{c2}^{HW}\right) n_{F}\left[ 1-\left( n_{c2}^{HW}/n_{F}\right) %
\right] ^{1/2}$. \ Furthermore, the asymptotic ( $n_{F}\rightarrow \infty $)
slope of $\overline{\Delta }_{GL}\left( n_{F}\right) $ for the best fitting
parameters is found to be very close to the corresponding HW slope, that is: 
$n_{c2}/\gamma \simeq n_{c2}^{HW}/\gamma _{HW}=\left( \mu /\hbar \omega
_{D}\right) \sinh \left( 1/\lambda \right) $ (see Fig.3). \ Thus, both $%
\Delta _{GL}\left( n_{F}\right) $ and $\Delta _{HW}\left( n_{F}\right) $
tend in the zero field limit to the well known result for the SC energy gap, 
$\Delta =\hbar \omega _{D}/\sinh \left( 1/\lambda \right) $. \ On the other
hand, for $n_{F}$ values below $n_{c2}$ (i.e. for $H>H_{c2}=\mu m^{\ast
}c/\hbar en_{c2}$), where $\Delta _{GL}\left( n_{F}\right) =0$, $\Delta
_{SC}\left( n_{F}\right) $ deviates dramatically from the semiclassical
theory, showing nonvanishing values, of magnitude comparable, or smaller
than $\hbar \omega _{c}$. The novel quantum mixed SC state created in this
field range is characterized by a cascade of normal to SC phase transitions,
which crossovers to the usual SC state with a monotonically, steeply
increasing order parameter amplitude for increasing $n_{F}$ (decreasing $H$
) values around $n_{c2}$ ($H_{c2}$). This crossover closely follows the
crossover of the quasi-particle spectrum from a well-defined Landau bands
structure to a continuum, as can be seen in Fig.(3), where the width, $%
\Delta n_{F}$, of the Landau band at the chemical potential, defined by the
interval of $n_{F}$ values corresponding to nontrivial self-consistent
solutions, is plotted as a function of integer $n_{F}$, together with $%
\Delta _{SC}\left( n_{F}\right) $. \ It is remarkable that, in the entire
fields range investigated, our calculations show that $\Delta n_{F}\simeq
0.2\Delta _{SC}\left( n_{F}\right) $, indicating that {MQ oscillations},
observed in the normal state, do not suffer appreciable additional damping
upon entering the quantum mixed SC state. Significant effect of the SC order
parameter on the {MQ} oscillations can be therefore observed only in the
crossover region, as found experimentally in the organic superconductor $%
\kappa -\left( ET\right) _{2}Cu\left( SCN\right) _{2}$ \cite{van der Wel95},%
\cite{Sasaki98},\cite{Maniv01}. The filling-factors range of this novel,
quantum mixed SC state, estimated by $n_{c2}-n_{c2}^{HW}\approx
n_{c2}^{HW}=0.563\left( \mu /\hbar \omega _{D}\right) ^{2}\sinh ^{2}\left(
1/\lambda \right) $, increases significantly (i.e. quadratically) with
decreasing values of $\hbar \omega _{D}/\mu $ , and much more sharply with
decreasing values of $\lambda $ \ (see also Fig.2).

Experimental evidence for the existence of the quantum mixed SC state
discussed above {can} be found in results reported for the high field
surface superconductivity observed recently in the topological insulator Sb$%
_{2}$Te$_{3}$ \cite{Zhao15}. Using a simple s-wave BCS model of a 2D
(circular) Fermi surface with the experimentally observed {dHvA frequency,} $%
F=36.5$ T, {and cyclotron mass} $m^{\ast }=0.065m_{e}$, {and employing the
basic dimensionless constants} $\lambda $ {and} $\hbar \omega _{D}/\mu $ of {%
the model as adjustable parameters, we fit the zero field limit of the
calculated self-consistent order parameter:} $\Delta _{SC}\left(
n_{F}\rightarrow \infty \right) \rightarrow \hbar \omega _{D}/\sinh \left(
1/\lambda \right) $ {to the average SC energy gap derived from the STS
measurements} (i.e. $\simeq 13$ meV) \cite{Zhao15}, and the semiclassical
critical field ( $n_{c2}^{HW}\approx 14$, see Fig.(3)) to the experimentally
determined field of the resistivity onset downshift $H_{R}$ ($\sim 2.5$ T, $%
n_{F}\sim 14$)\cite{Zhao15}. {The resulting values of the adjustable
parameters}, $\lambda =1$ and $\hbar \omega _{D}/\mu =0.25$, {imply strong
coupling superconductivity with relatively small cut-off energy for the
surface state of} Sb$_{2}$Te$_{3}$.

Now, using the set of parameters selected above, we've calculated the
self-consistent order parameter amplitude, $\Delta _{SC}\left( n_{F}\right) $%
, at the temperature of the experiment, $T=1.9$ K ( $k_{B}T/\mu =2.5\times
10^{-3}$), with the best fitting GL parameters; $\gamma =5.8$, and $%
n_{c2}=27 $, showing an extended quantum mixed state region above the
crossover field ($n_{c2}=27$, see Fig.(3)), characterized by small
Landau-band width, $\Delta n_{F}\lesssim 0.1$, which seems to account for
the puzzled, virtually normal state damping of the dHvA oscillation measured
in this system below $H_{R}$\cite{Zhao15}. \ 

{It should be noted that the ideal model system considered here (by ignoring
effects of disorder) is aimed at emphasizing the fundamental nature of our
predictions, which could be clearly observed only in sufficiently pure
materials. Nevertheless, the experimental results reported in Ref.}\cite%
{Zhao15}{\ indicate that for the 2D superconductivity, realized in
electronic surface states with Fermi energy near a Dirac point, the
stringent purity conditions are dramatically relaxed due to the drastic
suppression of the cyclotron mass. Thus, with} $m^{\ast }=0.065m_{e}$, {the
LL spacing of} $5$ {meV at} $H=3$ T, {exceeds the disorder scattering
relaxation} {rate (}$\sim 3$ {meV),} {observed in} Sb$_{2}$Te$_{3}$.{The
large scale of the cyclotron energy can also justify the neglect of the
Pauli limiting effect in our model, originating in the relatively large
spin-orbit splitting, estimated to be about }$1$ meV {in} Sb$_{2}$Te$_{3}$. {%
The latter remains smaller than} $\Delta n_{F}\simeq 0.2\Delta _{SC}\left(
n_{F}\right) $ {shown in Fig.3, even in the broad quantum multi-critical
region.}

\begin{figure}[tbp]
\label{fig3} \includegraphics[width=3.3in]{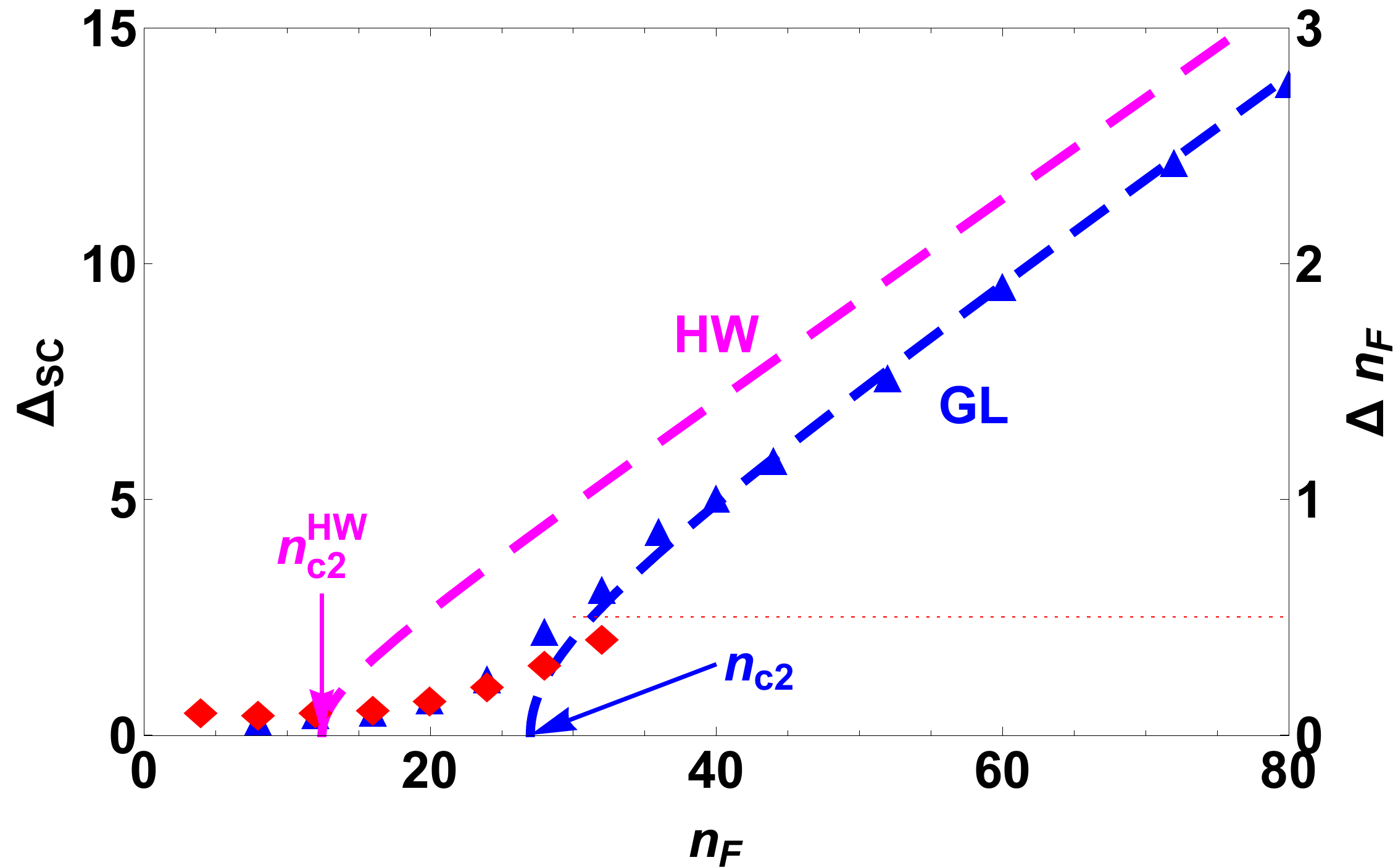}
\caption{Color online: Self-consistent order-parameter amplitde, $\overline{%
\Delta }_{SC}\left( n_{F}\right) $, calculated at integer values of $n_{F}$,
for $\protect\lambda =1.0$ , $\hbar \protect\omega _{D}=0.25\protect\mu $ ,
and $k_{B}T=2.5\times 10^{-3}\protect\mu $ (blue triangles), and the best
fitting curve, based on the GL-like expression 
\index{Del\_GL} (blue dashed line). Also shown is the corresponding
semiclassical (HW) result ( magenta dashed line) and the self-consistent
Landau bandwidth $\Delta n_{F}$ (red diamonds). The scale for $\Delta n_{F}$
on the right hand vertical axis. The horizontal red dotted segment indicates
the saturation value ($0.5$) of $\Delta n_{F}$ (for which the bands start to
overlap).}
\end{figure}

In conclusion, we have shown here that in 2D electron systems {with small
cyclotron mass at the Fermi energy}, where the effect of the magnetic field,
through Landau quantization, is most pronounced, self consistency of the SC
order parameter is crucial for understanding the transition to strong
type-II superconductivity. In particular, the single critical-point picture
of the SC phase transition, predicted by the mean-field semiclassical
theory, is smeared into a finite range multi-critical transition state,
characterized by a well defined Landau-bands structure in the quasi-particle
spectrum and suppressed SC order parameter amplitude. Upon decreasing
magnetic field below the semiclassical H$_{c2}$ the quasi particle spectrum
crossovers sharply into a continuum and the order-parameter amplitude
steeply approaches the well-known result predicted by the semi-classical
(Helfand-Werthamer) theory. The fields range of this quantum mixed\ SC phase
dramatically increases upon decreasing the pairing coupling constant $%
\lambda $, or cutoff energy $\hbar \omega _{D}$. \ It is therefore expected
that observable MQ oscillations can be significantly affected by the SC
order parameter only in the crossover region. This seems to be the reason
for the rather sporadic appearance of this effect in the diverse literature
reporting on dHvA oscillations in the SC state.\cite{Maniv01}\cite{Janssen98}
\ Our theory accounts reasonably well for the 2D SC state observed recently
on the surface of the topological insulator Sb$_{2}$Te$_{3}$ \cite{Zhao15},
revealing a strong type-II superconductor with unusually low carrier density
and small cyclotron effective mass, which can be realized only in the strong
coupling ($\lambda \sim 1$) superconductor limit. {This unique situation is
due to the proximity of the Fermi energy to a Dirac point, which implies
that other materials in the emerging field of surface superconductivity,
with metallic surface states and Dirac dispersion law around the Fermi
energy, can show similar features. }

This research was supported by E. and J. Bishop research fund at Technion.
T.M. is indebted to Dingping Li and B. Rosenstein for helpful discussions.

\end{document}